\documentclass[12pt]{article}

\usepackage{amsmath,amsfonts}
\usepackage{amsthm}

\newcommand{\aaa}{\mathcal{A}}
\newcommand{\atlas}{\mathcal{C}}
\newcommand{\borderop}{\delta} 
\newcommand{\bra}[1]{\pmb{\langle}#1\pmb{|}}
\DeclareMathOperator{\cardinality}{card}
\newcommand{\causet}{\vec{F}}
\newcommand{\cfield}{\mathbf{C}}
\newcommand{\dff}{\scshape}
\newcommand{\eee}{\mathbf{e}}
\newcommand{\fff}{{\cal F}}
\newcommand{\hhh}{\mathcal{H}}
\newcommand{\ia}{\Omega}
\newcommand{\idest}{{\itshape ie}}
\newcommand{\kd}{\text{\large\texttt{d}}}
\newcommand{\ket}[1]{\pmb{|}#1\pmb{\rangle}}
\newcommand{\kkk}{\mathcal{K}}
\newcommand{\llll}[5]{\put(#1,#2){\line(#3,#4){#5}}}
\DeclareMathOperator*{\lspan}{span}
\newcommand{\mmmm}[8]{\multiput(#1,#2)(#3,#4){#5}{\line(
#6,#7){#8}}}
\newcommand{\om}{\omega}

\newcommand{\pp}{\pi}
\newcommand{\qauset}{\vec{\Omega}}
\newcommand{\rfield}{\mathbf{R}}
\newcommand{\rota}{\propto}
\DeclareMathOperator*{\sfproj}{Proj}
\newcommand{\sfvec}{\mathbf{\omega}}
\newcommand{\ttt}{\mathcal{T}}
\newcommand{\tttt}{\mathbf{T}}
\newcommand{\unitelement}{\mathbf{1}}
\newcommand{\vertex}{{\circle*{1}}}
\newcommand{\vvv}{V}

\newcommand{\iak}{\ia(\kkk)} 
\newcommand{\ketbra}[2]{\ket{#1}\bra{#2}} 
\newcommand{\pkk}{{\pp:\kkk'\to\kkk}}

\title{Algebraic description of spacetime foam}

\author{Ioannis Raptis\thanks{\rm EU Marie Curie Postdoctoral Research 
Fellow, Theoretical Physics Group, Blackett Laboratory, Imperial College of 
Science, Technology and Medicine, Prince Consort Road, South 
Kensington, London SW7 2BZ, UK; e-mail: 
i.raptis@ic.ac.uk} and 
Roman R. Zapatrin\thanks{\rm 
Quantum Information Group, ISI, Villa Gualino, V.le S.Severo 65, 
10133, Torino, Italy; e-mail:  
zapatrin@isiosf.isi.it (address for correspondence)}}

\date{11 July 2002}

\begin{document}

\maketitle

\begin{abstract}

A mathematical formalism for treating spacetime topology as 
a quantum observable is provided. We describe spacetime foam 
entirely in algebraic terms. To implement the correspondence 
principle we express the classical spacetime manifold of general 
relativity and the commutative coordinates of its events by means 
of appropriate limit constructions.

\end{abstract}

\section*{Physical Motivation}

In this paper we present an algebraic model of spacetime foam. The 
notion of spacetime foam has manifold and somewhat ambiguous 
meaning in the literature if only because the models vary. There is 
no unanimous agreement about what foam `really' pertains to mainly 
due to the fact that each of the mathematical models highlights 
different aspects of that concept. Here we use the term `foam' 
along the concrete but general lines originally introduced by 
Wheeler \cite{wheeler64} who intended to refer to a spacetime with 
a dynamically variable, because quantally fluctuating, topology.  

The basic intuition is that at quantum scales even the topology of 
spacetime is subject to dynamics and interference. This conception 
of foam is in glaring contrast with general relativity, the 
classical theory of gravity, where spacetime is fixed to a 
topological manifold once and forever so that the sole dynamical 
variable is a higher level structure, namely, the spacetime 
geometry. It seems theoretically lame and rather {\itshape ad hoc} 
to regard the geometry of spacetime as being a dynamical variable 
attribute that can in principle be measured ({\idest}, an observable 
property of spacetime), while 
at the same time to think of its topology as a structure {\itshape 
a priori} fixed by the theoretician, an inert ether-like absolute 
background that is not liable to experimental investigation thus 
effectively an unobservable theoretical entity \cite{einst24}.  
Especially in the quantum realm where everything seems to be in a 
state of dynamical flux and quantum superposition, it appears to be 
quite plausible that not only the metric but also the spacetime 
topology manifests such a `standard' quantum behaviour. Hence, the 
theoretical investigation of spacetime foam as conceived above 
appears to be part and parcel of our apparently never ending quest 
for a cogent quantum theory of spacetime structure and its dynamics 
({\idest}, quantum gravity proper).

So, we will treat the topology of spacetime as a {\em quantum 
observable}, that is, as a measurable dynamical property of 
spacetime that engages into coherent quantum superpositions. In 
keeping with the general operationalist approach to quantum theory 
as originally championed by Heisenberg, we give an entirely {\em 
algebraic} formulation of spacetime foam. As befits the innate 
reticularity of quantum systems, our algebraic models are 
combinatorial structures of strongly finitistic character.  Indeed, 
such so-called locally finite or `finitary' models seem to be 
suitably designed to undermine the non-operational and unphysical 
nature of the classical spacetime continuum of general relativity.  
Laying the non-dynamical character of the manifold model for spacetime as 
described in the opening paragraph aside for a moment, its gravely 
non-operational trait is mainly that we have no actual experience 
of a continuous infinity of events since we always seem to record a 
finite number of them in laboratories of finite size during 
experiments of finite duration.  At the same time, its 
characterisation above as being unphysical pertains to that we can 
in principle pack an uncountable infinity of events in a finite 
spacetime volume which is likely to be the main culprit for the 
unrenormalisable infinities that plague a quantum field theoresis 
of gravity on a spacetime manifold.  Moreover, in an algebraic 
modelling of such discrete combinatorial spacetimes, one naturally 
expects that the representations of the algebras involved will also 
be finite dimensional if only to abide to the principle of 
finiteness as much as possible throughout the construction of the 
theory. All in all, it seems appropriate to call such a finitary 
algebraic approach to dynamical quantum spacetime topology or foam 
`pragmatic' \cite{qst}.

\medskip 

Of course, if the theory is to qualify as being physically 
plausible, realistic and literally pragmatic, then the notion of topology, 
commonly understood as the mathematical theory of the properties of 
space proper, seems an inappropriate starting point for the 
unification of relativity and the quantum. From a relativistic 
point of view `spatiality' or `spacelikeness' is fundamentally an 
unphysical conception of the connections between events since we 
have no experience of acausal or tachyonic dynamical propagations 
at least as regards our experimental familiarity with matter 
quantum field actions which seem to obey the principle of Einstein causality 
({\idest}, that acausally connected or spacelike separated 
matter quantum field operators (anti)commute). Thus topology, 
generally conceived as the study of properties of Euclidean, 
undirected spatial connections between points of space, seems to 
fare poorly against the inherently causal or temporal and directed 
nature of the connections between events that relativity theory 
brought to light \cite{zee64,bomb87,df88,rapt00a}. To emphasise the 
fundamentally causal or temporal, as opposed to spatial or 
topological proper, interpretation of our models, we may summarise 
our approach to the following: we study the dynamics and coherent 
quantum interference of causality, hence of the topology of 
space{\em time}, rather than merely of the topology of space. For 
this, it is perhaps more accurate to call the topology to be 
considered here {\itshape causal topology} and more appropriate to 
conceive our scheme as an algebraic description of some kind of 
{\em quantum causal foam}. 

Also from a quantum-theoretic point of view, the usual notion of 
topology seems rather inappropriate, for what does space(time) as 
an inactive pre-existent realm `out there'---one that is 
independent of the dynamical relations between quanta and our 
`measurement interactions' with them---mean in the 
quantum deep? To us the fundamental approach seems to be the other 
way around: it is the dynamical relations and interactions of 
quanta that {\em define} spacetime. In a more operationalistic note 
and in accord with the general philosophy of quantum mechanics that 
supports an observer dependent, or even created, physical reality, 
it is more likely that our own observations or measurements of the 
dynamics of quantum causality yields, in some conception and 
concomitant modelling of the process of measurement in the manner 
of Bohr's correspondence principle, the classical (and more 
familiar!) picture of spacetime as a topological manifold.  Indeed, 
for discretised quantum spacetime topologies this limit 
construction and its interpretation as the emergence of 
classicality {\itshape \`a la} Bohr, has already been proposed 
\cite{qst}; here we only reuse it in the context of fluctuating 
dynamical quantum causal topology. At least it has been contended 
that the commutative algebras of continuous ({\idest}, 
$C^{0}$) coordinates of the events of spacetime as a topological 
manifold arise precisely at such a classical limit \cite{rapt00b} 
with the latter being interpreted as a correspondence principle 
{\itshape \`a la} \cite{qst}.

\medskip 

Affine to the remarks above is the point that general 
relativity may be conceived as the dynamical theory of 
$g_{\mu\nu}(x)$, with $x$ living in a differential manifold $M$; 
hence, since the metric defines the local causal relations between 
events ({\idest}, $g_{\mu\nu}(x)$ delimits the lightcone at every 
event $x$ of $M$), gravity may be thought of as the dynamics of 
local causality---the dynamical tilting of the local lightcone due 
to the gravitational force.  In turn, it is customary in physical 
jargon to call local causality `locality', so that general 
relativity may be viewed as a sound dynamical theory of locality. 
The principle of locality in general relativity is successfully 
captured by the mathematical assumption that on top of $M$'s 
topological manifold ({\idest}, $C^{0}$) structure, there is also a 
differential geometric or $C^{\infty}$-smooth structure which, in 
turn, renders $g_{\mu\nu}(x)$ a smooth field. In $M$ causality 
connects and dynamically  evolves between infinitesimally separated 
events. That general relativity is a dynamical theory of the field 
of locality $g_{\mu\nu}$ is concisely encoded in the assumption of 
$M$ as a differential manifold. This $M$ seems to be the last 
classical vestige of an ether-like substance in our theories of 
spacetime structure and its dynamics \cite{einst24}---a classical 
relic that seems to be in irreconcilable discord with the 
inherently granular and algebraic character of the quantum 
\cite{einst36,einst56}.  Thus, it would be desirable that our 
algebraic and reticular models of quantum causal foam yield in the 
aforementioned limit not only the {\em topological}, but also the 
local {\em differential} structure of the classical spacetime 
manifold $M$ \cite{qst}. 

Related to these comments on locality is the relevant issue of 
spacetime event localisation. It has been amply justified that one 
cannot measure the local gravitational field $g_{\mu\nu}$ at an 
event $x$ with accuracy higher than Planck's space-time intervals 
$l_{P}\simeq 10^{-35}m$-$t_{P}\simeq 10^{-44}s$ without creating a 
black hole. This implies an innate granularity ({\idest}, a 
built in cut-off) and an irreducible fuzziness in operations of 
spacetime localization at quantum scales plus it questions the 
validity of the assumption of a geometrical point-set manifold for 
spacetime.  Indeed, it is well established that field theories like 
general relativity which have assumed a point-set spacetime 
continuum up-front are infested with singularities long before 
their quantisation becomes an issue.  This unpleasant phenomenon 
becomes even more acute in the quantum field theories of matter in 
the absence of gravity which suffer from infinities that are 
predominantly due to the point-like character of the sources of the 
fields involved, which sources, in turn, are thought of as 
occupying point events in the Minkowski continuum. All this seems 
to justify our primitive intuition of {\itshape ab initio} 
substituting or `smearing' the points of a manifold by larger or 
coarser open sets about them, which regions can also coherently 
superpose with each other---an intuition that lies at the heart of 
our approach to spacetime foam 
\cite{sork91,sork95,qst,rapt00b,mallrapt00}. This resembles the 
operator valued distribution approach to the flat quantum field 
theories of matter, as well as the by now standard method of 
`blowing up' singular points in algebraic geometry \cite{rapt00c}.  
To wrap things up, we claim that locality and localisation are 
purely classical conceptions intimately related to the manifold 
model of spacetime that in a quantum theoresis of spacetime 
structure and gravity must be abandoned only to be recovered in 
some classical limit. It follows that our algebraic structures 
modelling quantum causal foam are not only reticular, but also in 
quite a strong sense {\em alocal} \cite{qst}.

\medskip 

\noindent The present paper is organised as follows:

\begin{itemize} 
\item In section \ref{salgmod} we present the algebraic model for 
spacetime foam mainly concentrating on a special example of a 
method, quite standard in algebraic geometry, of extracting 
(finitary) topological spaces from (finite dimensional) algebras.
\item In section \ref{sclasscaus} we `causalize' our algebras in 
the sense that we give a sound causal interpretation to the 
extracted finitary topologies and then give a sound quantum 
interpretation to these locally finite causal topologies so that 
the epithet spacetime foam given to them is justified.  
\item In section \ref{sclcorr} we argue for a limit construction 
that recovers the classical spacetime continuum and the local 
differential structure of its events from an inverse system of 
these alocal dynamical quantum causal substrata.
\item In section \ref{stoymodel} a toy model giving us different 
spacetimes as values of a quantum observable is presented.  
\end{itemize} 

The afterword concluding the paper attempts to anticipate and reply 
to some `natural' criticism that one may exercise on our algebraic 
approach to spacetime foam, as well as it gives a brief account of 
what in our view will more likely be the future development of the theory.

\section{Algebraic model}\label{salgmod}

Let us now move to technicalities. Following a very general 
operationalist philosophy, with any (in particular, spacetime) 
measurement we associate an `operation device' as it is common 
practice in standard quantum mechanics; that is to say, a 
self-adjoint operator in an appropriate Hilbert space. The 
eigenvalues of the operator are the possible outcomes of acts of 
measurement on the quantum system in focus\footnote{In this work we 
take it almost axiomatically that {\em spacetime is a quantum 
system} \cite{qst}.}, and its eigenspaces are the `rooms' where the 
state of the system jumps after measurement. Note that, as a matter 
of fact, it is not relevant what particularly is drawn on the 
device's scale. This means that instead of a self-adjoint operator 
we can consider just a partition of the overall space into mutually 
orthogonal subspaces generated by an appropriate decomposition of 
the unit operator in the algebra of observables 
$\unitelement=\sum_{\alpha\in A}P_\alpha$. So, if we would have a 
procedure which associates with every summand in the 
above partition a topological 
space rather than a number, then we could say that topology has 
become a quantum observable. To be able to do that, we have to assume that 
the state space itself must possess some additional structure. 

According to our settings this additional structure is that of 
an {\em algebra}, associative but not commutative in general. In this 
section we describe the machinery for constructing topological 
spaces out of algebras which consists of: 

\begin{itemize}
\item Spatialisation procedure. Given a (finite dimensional) algebra, 
it extracts the set of points and endows it with the Rota topology. 
\item Geometrisation. We show that geometrical properties, such as 
the differential structure or `differentiability', are captured 
already at the finitary level, so that the $C^{\infty}$-smooth continuum 
is not a necessary prerequisite for their existence. 
\item Functoriality. We establish the duality between the algebras 
and appropriate geometries which enables us to build the limits 
which, in turn, are intended to support the classical correspondence 
principle. We address this issue in section \ref{sclcorr}. 
\end{itemize}

\subsection{Instead of manifolds}\label{ss21} 

We pass from the {\em raison d'\^etre} for reticular spacetimes 
presented in the previous section to their {\em fa\c con d'\^etre}.  
As it was claimed above, for every outcome of a pragmatic spacetime 
measurement we expect the result to be formulated in finitary 
terms. The idea to consider gravitational spacetime models 
consisting of finite number of points arises in Regge \cite{regge}, 
then at the level of topological spaces it was elaborated by Isham 
\cite{isham89} who suggested a lattice of finite topologies to 
serve as the (kinematical) configuration space for a canonical 
({\it ie}, Hamiltonian) topodynamics. To 
what extent 
can finite topological spaces be similar to manifolds? To give this 
question a precise meaning we recall the coarse-graining procedure 
or `algorithm' due to Sorkin \cite{sork91}. 

\paragraph{Finitary substitutes of topological spaces.} From a 
formal point of view and briefly, Sorkin's algorithm looks as 
follows. When we are speaking of spacetime as a topological or 
$C^{0}$-manifold $M$, its mere definition assumes that we have a 
covering or charting $\atlas$ of $M$ by open subsets.  The idea of 
coarse-graining is to replace the existing topology of the manifold 
$M$ by that generated by the covering $\atlas$. At the heart of 
this idea lies the primitive intuition that `large', `coarse' or `fuzzy' 
open sets are more `pragmatic' than points 
\cite{sork91,sork95,qst,buttish,rapt00b}, so that the geometric point-like 
continuum $M$ will be recovered at the ideal limit of infinite 
refinement of the $\atlas$s. As a result, the spacetime manifold 
acquires the cellular or simplicial structure with respect to 
$\atlas$, so that the events belonging to one cell are thought of 
as operationally indistinguishable. Then, instead of considering 
the set $M$ of all events we can focus on its quotient with respect 
to the relation 

\begin{equation}\label{etopeq}
x\equiv y \quad\hbox{ if and only if }\quad \forall {\cal O} \in 
\atlas
\quad x\in
{\cal O} \Leftrightarrow y\in {\cal O}
\end{equation} 

\noindent which is finite whenever the atlas $\atlas$ of $M$ is 
finite. 

Let us consider the behaviour of sequences of elements in finite 
topological spaces. First it is worth mentioning that if a finite 
topological space is Hausdorff, then its topology is necessarily 
discrete ({\idest}, in a sense degenerate). Since we are going 
to deal with non-trivial topologies, we should not expect them to 
be Hausdorff or $T_{2}$. As a consequence, the theorem of the 
uniqueness of the limit of a sequence will not be valid anymore.  
Another equivalent way of defining a topology $\tau$ on a set $X$ is 
to define which sequences in $X$ converge with respect to $\tau$ 
and which do not.  So, for finite $X$ we can instead of listing out 
or drawing pictures of open sets draw (or describe somehow) the 
graph of convergences of sequences $x,x,\ldots \, x,
\ldots\to y$, this indicating that the constant sequence $\{ x\}$ 
converges to $y$ in the topology $\tau$. Consider a couple of 
examples in which, to distinguish between open and closed 
intervals, we write 

\[
\begin{array}{rcl}
(a,b)&=&\{x:\,a<x<b\}\cr
[a,b]&=&\{x:\,a\le{}x\le{}b\}
\end{array}
\]

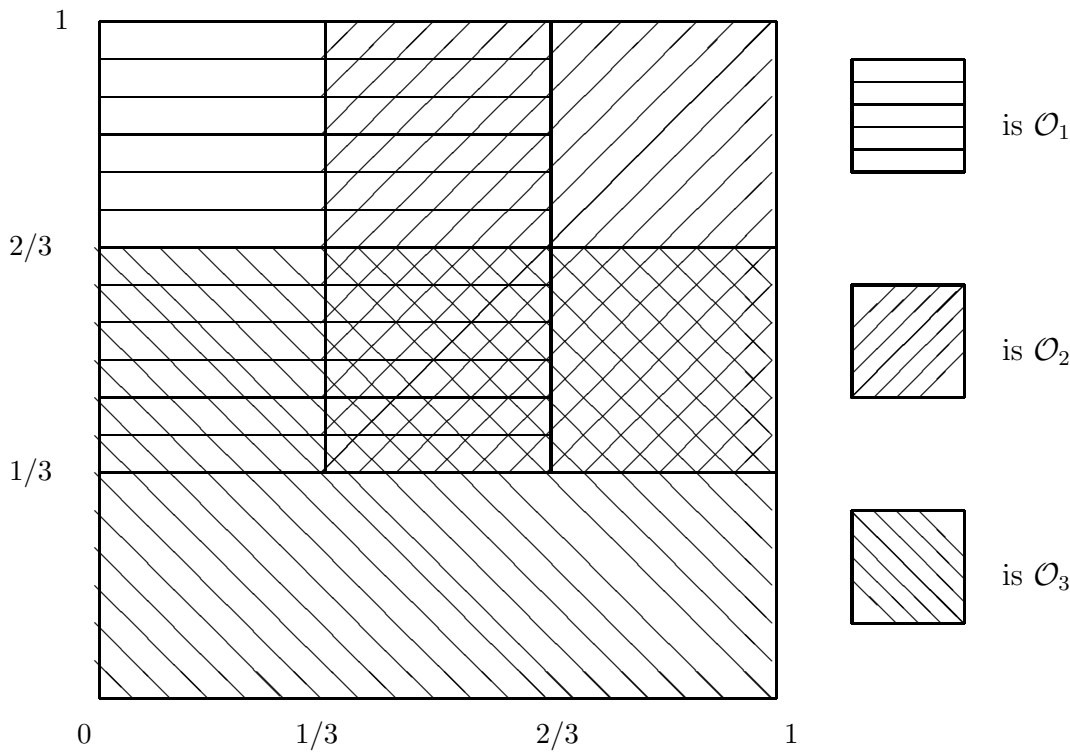
\begin{figure}[h!t]

\begin{center}

\begin{picture}(120,90)
\thicklines
\llll{0}{00}{1}{0}{90}
\llll{0}{30}{1}{0}{90}
\llll{0}{60}{1}{0}{90}
\llll{0}{90}{1}{0}{90}

\llll{0}{0}{0}{1}{90}
\llll{30}{30}{0}{1}{60}
\llll{60}{30}{0}{1}{60}
\llll{90}{0}{0}{1}{90}

\put(-3,-6){\mbox{\small 0}}
\put(26,-6){\mbox{\small 1/3}}
\put(58,-6){\mbox{\small 2/3}}
\put(91,-6){\mbox{\small 1}}
\put(-12,29){\mbox{\small 1/3}}
\put(-12,59){\mbox{\small 2/3}}
\put(-6,89){\mbox{\small 1}}

\llll{100}{70}{1}{0}{15}
\llll{100}{70}{0}{1}{15}
\llll{100}{85}{1}{0}{15}
\llll{115}{70}{0}{1}{15}
\put(120,75){\mbox{is ${\cal O}_1$}}

\llll{100}{40}{1}{0}{15}
\llll{100}{40}{0}{1}{15}
\llll{100}{55}{1}{0}{15}
\llll{115}{40}{0}{1}{15}
\put(120,45){\mbox{is ${\cal O}_2$}}

\llll{100}{10}{1}{0}{15}
\llll{100}{10}{0}{1}{15}
\llll{100}{25}{1}{0}{15}
\llll{115}{10}{0}{1}{15}
\put(120,15){\mbox{is ${\cal O}_3$}}

\thinlines

\mmmm{0}{35}{0}{5}{11}{1}{0}{60}

\newcounter{llg}
\setcounter{llg}{60}
\multiput(28,30)(0,5){12}{
\line(1,1){\value{llg}}
\addtocounter{llg}{-5}
}
\setcounter{llg}{60}
\multiput(28,30)(5,0){12}{
\line(1,1){\value{llg}}
\addtocounter{llg}{-5}
}

\setcounter{llg}{60}
\multiput(-2,60)(0,-5){12}{
\line(1,-1){\value{llg}}
\addtocounter{llg}{-5}
}
\multiput(3,60)(5,0){5}{\line(1,-1){60}}
\setcounter{llg}{60}
\multiput(28,60)(5,0){12}{
\line(1,-1){\value{llg}}
\addtocounter{llg}{-5}
}

\multiput(100,73)(0,3){4}{\line(1,0){15}}
\setcounter{llg}{15}
\multiput(98.5,40)(0,3){5}{
\line(1,1){\value{llg}}
\addtocounter{llg}{-3}
}
\setcounter{llg}{15}
\multiput(98.5,40)(3,0){5}{
\line(1,1){\value{llg}}
\addtocounter{llg}{-3}
}

\setcounter{llg}{15}
\multiput(98.5,25)(0,-3){5}{
\line(1,-1){\value{llg}}
\addtocounter{llg}{-3}
}
\setcounter{llg}{15}
\multiput(98.5,25)(3,0){5}{
\line(1,-1){\value{llg}}
\addtocounter{llg}{-3}
}

\end{picture}

\end{center}

\caption{The covering of a piece of plane.}

\label{fcovplane}
\end{figure}

\paragraph{Example 1.} Let $M$ be a piece of plane:  $M=(0,1)\times
(0,1)$, and $\atlas = \{{\cal O}_1, {\cal O}_2, {\cal O}_3\}$ be its
covering with ${\cal O}_1=(0,\frac{2}{3})\times (\frac{1}{3},1)$,
${\cal O}_2=(\frac{1}{3},1)\times (\frac{1}{3},1)$ 
and ${\cal O}_3=(0,1)\times(0,\frac{2}{3})$ (See Fig. 
\ref{fcovplane})

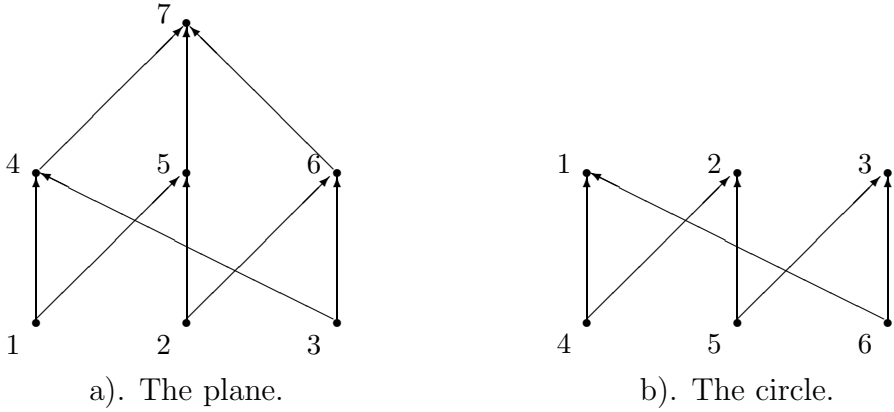
\begin{figure}[h!]
\begin{center}
\begin{tabular}{ccc}
\begin{picture}(40,40)
\multiput(0,0)(20,0){3}{{\vertex}}
\put(-4,-4){\mbox{\small 1}}
\put(16,-4){\mbox{\small 2}}
\put(36,-4){\mbox{\small 3}}
\multiput(0,20)(20,0){3}{{\vertex}}
\put(-4,20){\mbox{\small 4}}
\put(16,20){\mbox{\small 5}}
\put(36,20){\mbox{\small 6}}
\put(20,40){{\vertex}}
\put(16,40){\mbox{\small 7}}
\multiput(0,0.5)(20,0){3}{\vector(0,1){19}}
\multiput(0,0.5)(20,0){2}{\vector(1,1){19}}
\put(39.5,0.5){\vector(-2,1){39}}
\put(0.5,20.5){\vector(1,1){19}}
\put(20,20.5){\vector(0,1){19}}
\put(39.5,20.5){\vector(-1,1){19}}
\end{picture}
&$\qquad\qquad\qquad$& 
\begin{picture}(40,40)
\multiput(0,0)(20,0){3}{{\vertex}}
\put(-4,-4){\mbox{\small 4}}
\put(16,-4){\mbox{\small 5}}
\put(36,-4){\mbox{\small 6}}
\multiput(0,20)(20,0){3}{{\vertex}}
\put(-4,20){\mbox{\small 1}}
\put(16,20){\mbox{\small 2}}
\put(36,20){\mbox{\small 3}}
\multiput(0,0.5)(20,0){3}{\vector(0,1){19}}
\multiput(0,0.5)(20,0){2}{\vector(1,1){19}}
\put(39.5,0.5){\vector(-2,1){39}}
\end{picture}
\cr
&&
\cr
\mbox{a). The plane.}&&\mbox{b). The circle.}
\end{tabular}
\end{center}
\caption{The graphs of convergences for the finitary 
substitutes for the plane and the circle. The numbers label the 
vertices to concord with the matrix representations \eqref{exofalg}.} 
\label{fcircle} 
\end{figure}

Let us describe in detail how the convergence graph (Fig. 
\ref{fcircle}a) is built.  Denote by subsequent numbers the 
following subsets of $M$ 

\[
\begin{array}{r@{=}l@{\qquad}r@{=}l}
1&(0,\frac{1}{3})\times(\frac{2}{3},1)&
4&(0,\frac{1}{3})\times[\frac{1}{3},\frac{2}{3}]\cr
2&(\frac{1}{3},1)\times(\frac{2}{3},1)&
5&[\frac{1}{3},\frac{2}{3}]\times(\frac{2}{3},1)\cr
3&(0,1)\times(0,\frac{1}{3})&
6&(\frac{2}{3},1)\times[\frac{1}{3},\frac{2}{3}]\cr
\end{array}
\] 
\[
7=[\frac{1}{3},\frac{2}{3}]\times[\frac{1}{3},\frac{2}{3}]
\] 

\noindent which stand for the equivalence classes of the relation 
\eqref{etopeq}. 

The arrows drawn on Fig. \ref{fcircle}a appear as follows.  Take, 
say, a point in the set 1 and a point in the set 4.  Then any set 
from $\atlas$ containing 1 contains 4, therefore 1 tends to 4, 
which is depicted by drawing the appropriate arrow. All other 
arrows are obtained likewise. 

\paragraph{Example 2.} A circle. Let $M=\exp(i\phi)$, and let the 
covering be $\atlas = \{{\cal O}_1 ,{\cal O}_2, {\cal O}_3\}$ with 

\[
\begin{array}{l}
{\cal O}_1 = (-\pi/3,5\pi/6) \cr
{\cal O}_2 = (\pi/2,9\pi/6) \cr
{\cal O}_3 = (-2\pi/3,\pi/3) 
\end{array}
\]

\noindent for which the convergence graph is built in a similar way 
and shown on Fig. \ref{fcircle}b.

\medskip 

It is straightforward to prove that for any points $x,y,z,$ of a
finite topological space $X$ $x\to y$ and $y\to z$ imply $x\to
z$. That means that its graph of convergences, denoted by $X$, 
will be always {\em transitive}. This remarkable property of 
finitary substitutes makes it possible to associate algebras with 
them, called Rota algebras. 

\medskip 

\subsection{Rota algebras}\label{ssrotalg} 

\paragraph{Rota algebras in Dirac notation.} Let $X$ be a finite 
topological space. Consider the linear space $\ia$ whose basis 
$\ketbra{i}{j}$ is labelled by tending pairs $i\to j$ of points of 
$X$. 

\begin{equation}\label{edefrotalg}
\ia(X)=
\left\{
\sum_{i,j\in X}\limits\ketbra{i}{j}
\quad\mbox{such that}\quad i\to j
\right\}
\end{equation}

\noindent In the sequel, when no confusion occurs, we omit the 
notation of the topological space $X$ in parentheses and simply 
write

\[
\ia=\ia(X)
\]
 
\noindent Define the product on $\ia$ by setting it on its basic 
elements:

\begin{equation}\label{edef22}
\ketbra{i}{j}\ketbra{k}{l}\quad =\quad
\left\lbrace\begin{array}{lcl}
\ketbra{i}{l} &,& \hbox{if} \quad j=k \cr
0 &,& \hbox{otherwise}
\end{array}\right.
\end{equation}

\noindent Note that $\ketbra{i}{l}$ in \eqref{edef22} is always 
well-defined since the relation of convergence ``$\to$" is always 
transitive, that is why the existence of darts $i\to j$ and $j\to 
k$ in the convergence graph of $X$ always enables the existence of 
$i\to k$.  The space $\ia$ with the product \eqref{edef22} is 
called the {\dff Rota algebra} of the topological space $(X,\to)$. 
These algebras were first introduced in \cite{rota} in the context 
of combinatorial theory. 

\paragraph{The matrix representation of Rota algebras.} Given the 
Rota algebra of a finite topological space $X$, its standard matrix 
representation is obtained by choosing the basis of $\ia$ 
consisting of the elements of the form $\ketbra{i}{k}=\eee_{ik}$, 
with $ik$ ranging over all converging pairs $i\to k$ of elements of 
$X$. The matrices $\eee_{ik}$ (called matrix units) are 
defined as follows:

\begin{equation}\label{eab}
\eee_{ik}(m,n)= 
\left\lbrace\begin{array}{rl}
1 & \mbox{$m=i$ and $n=k$ (provided $i\to k$)}\cr
0 & \mbox{otherwise}
\end{array}\right.
\end{equation}

\noindent where $\eee_{ik}(m,n)$ stands for the element in the 
$m$-th row and the $n$-th column of the matrix $\eee_{ik}$. We can 
also extend the ranging to {\em all} pairs of elements of $X$ by 
putting $\eee_{ik}\equiv 0$ for $i\not\to k$. Then the product 
\eqref{edef22} reads:

\begin{equation}\label{mprod}
\eee_{ik}\eee_{i'k'} = \delta_{ki'}\eee_{ik'}
\end{equation}

\medskip 

To specify a Rota algebra in the standard matrix representation we 
fix the template matrix replacing the unit entries in the incidence 
matrix $I_{ik}$ of the graph $X$: 

\[ 
I_{ik} = 
\left\lbrace\begin{array}{rl}
1 & i\to k\cr
0 & \mbox{otherwise}
\end{array}\right.
\]

\noindent by wildcards $*$ ranging independently over all numbers.  
For the examples considered above the templates for their Rota 
algebras have the following form: 

\begin{equation}\label{exofalg}
\ia({\hbox{\small plane}}) = 
\left(\begin{array}{ccccccc}
*&0&0& *&*&0& *\cr
0&*&0& 0&*&*& *\cr
0&0&*& *&0&*& *\cr
0&0&0& *&0&0& *\cr
0&0&0& 0&*&0& *\cr
0&0&0& 0&0&*& *\cr
0&0&0& 0&0&0& *
\end{array}\right)
\quad ;\quad
\ia({\hbox{\small circle}}) = 
\left(\begin{array}{cccccc}
*&0&0& 0&0&0\cr
0&*&0& 0&0&0\cr
0&0&*& 0&0&0\cr
*&0&*& *&0&0\cr
*&*&0& 0&*&0\cr
0&*&*& 0&0&*
\end{array}\right)
\end{equation}

\medskip

\noindent where the wildcard $*$ denotes the ranging over the field
of numbers; for instance, the algebra associated with the two-point space with 
the topology 
\(\unitlength=0.2ex
\begin{picture}(23,4)
\put(0,4){\circle*{3}}
\put(20,4){\circle*{3}}
\put(3,4){\vector(1,0){15}}
\end{picture}
\) 
has the the following template matrix 

\[
\ia({\mbox{\(\unitlength=0.1ex
\begin{picture}(23,4)
\put(0,4){\circle*{3}}
\put(20,4){\circle*{3}}
\put(3,4){\vector(1,0){15}}
\end{picture}
\)}})= 
\left( \begin{array}{cc}
* & * \\
0 & *
\end{array} \right)
\,=\,
\left\lbrace \left.
\left( \begin{array}{cc}
a & b \\
0 & c
\end{array} \right)
\right\vert
\;
a,b,c \in \cfield
\right\rbrace
\]

\medskip 

So, we see that any finite topological space can be described in 
terms of a finite-dimensional algebra (for more details we refer to 
\cite{fas}).

\subsection{Spatialisation and Rota topology}\label{ssrotatop}

Here we show how to associate a finite topological space with an 
arbitrary finite-dimensional algebra. 

\paragraph{The emergence of points.} Let us start with a given 
finite-dimensional associative (and non-commutative, in general) 
algebra $\ia$. According to standard conceptions and methods of 
modern algebraic geometry, as well as the general algebraic 
approach to physics, we assume points to be irreducible 
representations (IRs) of $\ia$. Let us dwell on this issue in a bit 
more detail. All the contents of our paper respects the so-called 
`observability principle'. This means that the spatialisation ({\em 
viz.} the emergence of points) should be considered with respect to 
the measurements we possess. This, in turn, is supposed to be 
captured by the notion of an algebra $\ia$ of observables available 
for experimental discourse with the quantum system in focus (here, 
spacetime). Starting from this algebra we have to introduce the 
notion of points as something to be regarded as `undivisible', or 
`inseparable', or even `atomic'---an irreducible `{\itshape 
ur-element}' from a set-theoretic or purely geometric point of 
view. From an algebraic perspective, as it is generally assumed,  
irreducible representations (IRs) of the algebra of observables 
involved seem to meet best this `elementarity of points' 
requirement. So, the first step of the spatialisation procedure is 
creating (or finding) points:  

\begin{equation}\label{eptsirs}
\{\,\mbox{points}\,\} =
\{\,\mbox{IRs}\,\}
\end{equation}

\paragraph{Standard set with nonstandard topology.} When the first 
spatialization step in a standard way \eqref{eptsirs} is done we may 
wish to proceed by endowing the set of points with a topology.  
There are standard recipes for this step as well like, say, the 
Zariski topology on the prime spectrum of $\ia$. Unfortunately, on 
finite-dimensional algebras this topology is always discrete, which 
leaves us no chance to fit the above requirement of being 
non-Hausdorff ({\idest}, not $T_{2}$). So, we are compelled 
to find another topology. 

\medskip 

As it was mentioned above, to define a topology on a finite set, it 
is enough to define it in terms of convergences, namely, to state 
which point tends to which \cite{ishamnato}. These terms are the 
most appropriate for describing the Rota topology (first it was 
introduced in \cite{fas}). 

Let $\ia$ be a finite-dimensional algebra. Denote by $X$ the set of 
points of $\ia$, each of which we shall associate with a prime 
ideal in $\ia$. Consider two points (representations of $\ia$)  
$x,y\in X$ and denote by the same symbols $x,y$ their kernels as 
it produces no further confusion. Both of them, being kernels of 
representations, are two-sided ideals in $\ia$, in particular,  
subsets of $\ia$, hence both of the following expressions make 
sense:

\[
x\cap y \subset \ia 
\qquad\mbox{and}\qquad
x\cdot y \subset \ia
\] 

\noindent the latter denoting the product of subsets of $\ia$: 
$x\cdot y=\{a\in \ia\mid\, \exists u\in x,\, v\in y:\, uv=a\}$. 
Since $x,y$ are ideals, we always have the inclusion $x\cdot 
y\subseteq x\cap y$ which may be strict or not. Define the relation 
$\rota$ on $X$ as follows: 

\begin{equation}\label{edefrota}
x\rota y\quad
\mbox{if and only if} \quad
xy\neq x\cap y
\end{equation}

\medskip 

Then the {\dff Rota topology} is the weakest one in which $x\rota 
y$ implies the convergence $x\to y$ of the point $x$ to the point 
$y$.  Explicitly, the necessary and sufficient conditions for $x$ 
to converge to $y$ in the Rota topology reads: 

\begin{equation}\label{eqxy}
x\to y
\quad\mbox{if and only if}\quad
\exists y_0,\ldots,y_n\,\mid\,
y_0=x,\,y_n=y;\,y_{k-1}\rota y_k
\end{equation}

\noindent This operation is called the transitive closure of the 
relation $\rota$. Note that, in general, the Rota topology can be 
defined on any set of ideals.

\medskip 

It was proved by Stanley \cite{stanley} that in that particular 
case when $\ia$ is the Rota algebra of a finite topological space, 
then its spatialisation $X$ endowed with the Rota topology 
\eqref{eqxy} is homeomorphic to the initial topological space. 
However, in general if we have two finite topological spaces and a 
continuous mapping between them, their Rota algebras will {\em not} 
be homomorphic. Recently, `good' classes of topological spaces were 
discovered for which the transition to Rota algebras is functorial 
\cite{dgreech,iasc}. In our approach, it supports the existence of 
the classical limit. We return to this issue in section 
\ref{sclcorr}. 

\subsection{Graded structures and geometrisation}\label{ssgraded} 

Here we discover that the topological spaces which arise 
in Sorkin's finitary substitution procedure are typically `good' 
from the point of view of their Rota algebras. We show that the 
finite-dimensional analogues of differential calculi can be 
introduced on finite topological spaces as well. 

\paragraph{Basic definitions.} The `good' topological spaces we 
mentioned above are abstract simplicial complexes. They are 
defined as follows. Let $\vvv$ be a non-empty finite set. Call the 
elements of $\vvv$ vertices. A collection $\kkk$ of non-empty 
subsets of $\vvv$ is called (abstract) {\dff simplicial complex} 
with the set of vertices $\vvv$ whenever

\begin{itemize}
\item $\forall v\in\vvv \quad \{v\}\in\kkk$
\item $\forall P\in\kkk,\, \forall Q\subseteq\vvv \quad
Q\subseteq P\Rightarrow Q\in\kkk$
\end{itemize}

\medskip 

The elements of $\kkk$ are called {\dff simplices}. The topology on 
$\kkk$ is defined in terms of convergences: 

\[ 
P\to Q 
\quad\stackrel{\mbox{Def}}{\Leftrightarrow}\quad
P\subseteq Q
\] 

\noindent where the latter inclusion makes sense since $P,Q$ are 
subsets of $V$. 

\medskip 

Now let $M$ be a topological space and $\atlas$ an arbitrary 
covering of it (call the elements of $\atlas$ charts). With any 
such $\atlas$ a simplicial complex can be associated according to 
the following definion. A {\dff nerve} of the covering $\atlas$ is 
a simplicial complex $\kkk(\atlas)$\footnote{This so-called 
`nerve-construction' of simplicial complexes from (open) covers of 
a topological space is originally due to \v{C}ech \cite{eilsteen}; 
see also \cite{al56}. We wish to thank the referee of Classical and 
Quantum Gravity for bringing to our attention this fact.} such that 

\begin{equation}\label{edefnerv}
\begin{array}{l} 
V=\atlas
\cr
P\in\kkk\;\Leftrightarrow\;\cap P\neq\emptyset
\end{array} 
\end{equation} 

\noindent Here the first condition \eqref{edefnerv} means that the 
vertices of $\kkk$ are associated with the charts of the covering 
$\atlas$, and the second condition \eqref{edefnerv} means that the 
simplices $P$ of $\kkk$ are the sets of charts of $\atlas$ whose 
intersection is non-empty.

\medskip 

So, given a covering of $M$, we can in principle form two finitary 
topological spaces out of it: the finitary substitute $X$ described 
in section \ref{ss21} and the nerve $\kkk$ of $\atlas$. In the case 
when the covering 

\begin{itemize}
\item $\atlas$ is minimal ({\idest}, when we cannot delete any of 
its elements for $\atlas$ to remain a covering) and
\item generic ({\idest}, when all nonempty intersections are different), 
\end{itemize}

\noindent these two topological spaces are the same. To show it, 
note that the minimality enables the first condition 
\eqref{edefnerv}. Then with any equivalence class \eqref{etopeq}
of points of $M$ we associate  the following simplex $P\in\kkk$: 

\[
P=\left\{A\in\atlas\mid\;
P\subseteq A\right\}
\] 

\noindent and the second condition \eqref{edefnerv} for $\atlas$ to 
be generic enables this correspondence to be one-to-one. 

\medskip 

From now on we assume that finite topological spaces $X$ we are 
dealing with are simplicial complexes. Let us study the structure 
of their Rota algebras. First note that $X$ is a graded set, 
namely, with any simplex its {\dff dimension} is associated: 

\begin{equation}\label{e53}
\dim P = \cardinality P - 1
\end{equation}

\noindent ---the number of its vertices minus one. 

\noindent Denote by $\kkk^n$ the $n$-skeleton of $\kkk$---the 
set of its simplices of dimension $n$, 
$\kkk^n=\{P\in\kkk:\:\dim P = n\}$ and consider the linear spans

\[
\hhh^n = \lspan\kkk^n =
\left\{\sum_{P\in\kkk^n}c_P\ket{P}\right\}
\]

\medskip 

The {\dff border operator} $\borderop:\hhh^n\to\hhh^{n-1}$ 
acts as

\begin{equation}\label{e53b}
\borderop\ket{P} = \sum_{v\in P}\epsilon_{vP}\ket{P_v}
\end{equation}

\noindent (where $\epsilon_{vP}=\pm1$ are the appropriate incidence 
coefficients) and then extends to the space $\hhh$ (assuming 
$\borderop\ket{v}=0,\,\forall v\in\vvv$):

\[
\hhh = \oplus \hhh^n = \lspan\kkk
\]

\medskip 

\noindent Taking advantage of the Dirac notation, the same symbol 
$\borderop$ can be used to denote its adjoint, called the {\dff 
coborder operator}, which acts from $\hhh^{*n}$ to $\hhh^{*n+1}$, as 
follows

\begin{equation}\label{e54e}
\forall P\in\kkk^n \qquad
\left\lbrace\begin{array}{rcl}
\bra{P}\borderop &\in& \hhh^{*n+1}\\
\borderop\ket{P} &\in& \hhh^{n-1}
\end{array}
\right.
\end{equation}

\medskip 

The Rota algebra $\ia(\kkk)$ defined in \eqref{edefrotalg} also has 
a graded structure, namely with each basic element $\ketbra{P}{Q}$ 
the number 

\[
\mbox{deg}\ketbra{P}{Q} = 
\dim Q - \dim P
\] 

\noindent is associated. It follows directly from the 
multiplication rule \eqref{edef22} that $\mbox{deg}$ is really a grading. 
Thus, $\ia(\kkk)$ can be written as a direct sum of its graded 
subspaces in the following manner

\begin{equation}\nonumber
\ia(\kkk)=\bigoplus_{i}\ia^{i};~~\ia^{i}:={\rm 
span}_{\cfield}\{\ketbra{P}{Q}:~\mbox{deg}\ketbra{P}{Q}=i\}
\end{equation}  

\paragraph{The differential structure on $\iak$.} Let us show that 
$\iak$ always has the structure of a differential module over 
$\aaa$, namely, that induced by the projection $\sigma$ onto the 
quotient. Now let us explicitly provide the form of the 
differential $\kd$ on $\iak$.  The following theorem holds (for the 
proof we refer to \cite{iasc}). 

\begin{quote}
\noindent The differential in the incidence algebra $\iak$ of a 
simplicial complex $\kkk$ has the following form. Let 
$\ketbra{P}{Q}\in\ia^n$, then

\begin{equation}\label{e62}
\kd\ketbra{P}{Q}=
\ket{\borderop P}\bra{Q}-
(-1)^n\ket{P}\bra{Q\borderop}
\end{equation}

\noindent where $\borderop$ is the symbol 
\eqref{e53b}--\eqref{e54e} for both border and coborder operations.

\end{quote}

\medskip 

So, we have established that `good' coverings capture not only 
the $C^{0}$-topology (as it was shown by Sorkin 
\cite{sork91}, see also \cite{rapt00b}), but they also 
possess the differential structure similar to the module of 
differential forms. Namely, 

\begin{itemize}
\item The r\^ole of the algebra of smooth functions is played by the 
diagonal part (or, in other words, of the component of degree 
$0$)---denote it $\aaa$---of its Rota algebra. \[ \aaa \equiv \ia^0 \] 
\item The $k$-degree component $\ia^k$ of the Rota algebra plays 
the r\^ole ({\em ie}, has the same modular properties with respect 
to the algebra $\aaa$) as the module of smooth differential forms 
of order $k$.  \item The operator $\kd$ \eqref{e62} is the 
finite-dimensional analog of the Cartan-K\"ahler differential.  
\end{itemize}

\section{Classical and quantum causality}\label{sclasscaus}

As promised in the physical motivation opening the paper, in this 
section we intend to replace the finitary topologies described above by 
mathematically  equivalent structures whose physical interpretation 
however is distinctly {\em chrono}-logical, or more generally, {\em 
causal}, rather than simply {\em topo}-logical or {\em spatial}.  
Then we will `algebraicise' the resulting causal spaces {\itshape 
\`a la} Rota, thus arrive swiftly at structures that may be soundly 
interpreted as {\em quantum causal spaces} supporting quantum 
causal topologies. Thus we will possess a concrete theoretical 
({\idest}, structural-mathematical) paradigm of {\em quantum causal 
foam}.   

\paragraph{Classical causal sets.} First, we recall the `classical' 
causal sets from \cite{bomb87} (the epithet `classical' for 
the causal sets of Bombelli {\itshape et al.} will be justified 
shortly). An `abstract' {\dff causal set} $\causet$ is a locally 
finite partially ordered set (poset) whose partial order is 
physically interpreted as the causal `after' relation between its 
event-vertices. Local finiteness pertains to the technical trait 
that every Alexandrov interval $A(x,y):=\{ z:~ x\to z\to y\}$ in 
$\causet$ is finite.  This must be compared with the `finitarity' 
property of the posets modelling 
finite topological spaces obtained from locally finite open 
coverings $\atlas$ of a bounded region $X$ of a spacetime 
topological manifold $M$ by Sorkin's algorithm \cite{sork91} 
as it was briefly discussed in \ref{ss21}. One easily verifies that the 
latter posets are {\itshape a fortiori} locally finite in the 
causal sets' sense.  Thus, one may consider a region $X$ bounded 
both in space {\em and} in time in a (possibly curved) spacetime 
manifold $M$ motivated mainly by the `pragmatic' fact that all our 
spacetime experiments are of finite spatial extension and temporal 
duration, as well as that in the course of them we record a finite 
number of events \cite{qst}. 

The causal reinterpretation of the posets in \cite{sork91} {\em 
from topological to causal} rests on the assumption that the open 
covers of $X$ now represent coarse (because dynamically and 
quantally perturbing)  approximations (`observations' or 
`measurements') of the causal relations between the events of $X$ 
rather than between their spatial topological relations proper 
\cite{rapt00a,mallrapt00}. Convincing arguments supporting such a 
reinterpretation of the relevant poset structures from topological 
to causal were originally given in \cite{sork95}.  Subsequently in 
\cite{rapt00a}, this reinterpretation was coined {\em 
causalisation}. Causalisation was somehow forced on us by the 
theory of relativity which mandates that the {\em physical} 
connections between events are the causal ones, not the tachyonic, 
spatial or space-like ones \cite{df88}.

\paragraph{Quantisation of causality.} Of course, if our 
mathematical models of finitary causal spaces were merely `arrow 
structures' ({\idest}, essentially small poset categories or 
arrow semigroups, or even monoids) they would only allow for serial 
concatenations of causal arrows to form causal paths, but would not 
be able to model coherent superpositions of them. To actually model 
{\em quantum} interferences of causality we would have to pass from 
poset to algebra structures. Indeed, if we were able to somehow 
associate algebras with the locally finite posets modelling causal 
sets and interpret their formal `$+$' operation as coherent quantum 
superposition of the latter's causal  arrows, we would arrive 
straightforwardly to structures which can be interpreted as quantum 
causal sets in much the same way that Sorkin's finitary topological 
spaces \cite{sork91} were algebraicised and concomitantly quantised 
in \cite{qst}. We do it by associating Rota algebras with 
Bombelli's {\itshape et al.} causal sets like when we associated 
incidence algebras to finitary topological posets in section 
\ref{salgmod}, as well as to study their finite dimensional 
irreducible representations (IRs) as befits quantum systems with 
finitely many degrees of freedom. The passage from a locally finite 
poset modelling a `classical' causal set $\causet$ to its 
associated incidence Rota algebra \eqref{edefrotalg}

\[ \qauset=\ia \left( \causet \right) \]

\noindent treated as {\dff quantum causal set} $\qauset$ 
introduced in \cite{rapt00a} was coined {\em `quantisation of 
causality'}. As it was also mentioned in connection 
with \eqref{edef22} in \ref{ssrotalg}, the transitivity of the 
basic partial order relation defining a causal set enables us to 
define an associative product structure in its associated incidence 
algebra ({\idest}, quantum causal set). 

It must be stressed however that the aforementioned algebraisation 
of finitary causality achieved by passing from the non-interfering 
causal sets of Sorkin {\it et al.} to the quantally superposing 
elements of their associated incidence Rota algebras modelling 
quantum causal sets is by no means the sole `justification' for 
studying the latter in the context of dynamically variable quantum 
causal topology and, {\itshape in extenso}, quantum gravity.  
Another issue of crucial importance is our ability to regard such 
structures as sound models of local causality---the potent 
quantal aspects of such algebraic models aside. That is, the 
principal (perhaps the sole!) dynamical variable in the quantum 
spacetime deep is {\em local quantum causality}. 

In an inherently finitistic and algebraic setting like ours 
locality cannot afford a meaning similar to the one it connotes in 
the context of the classical differential geometric continuum model 
for spacetime adopted exclusively as a realm in which the 
differential equations for gravity may be meaningfully (in fact, 
uniquely!) written down.  Instead of conceiving of gravitational 
actions as being represented by a $C^{\infty}$-smooth field of 
`local causality' or `locality', $g_{\mu\nu}(x)$, dynamically 
evolving between infinitesimally separated events in a differential 
manifold spacetime $M$, one may think of causal influences in a 
reticular `environment' that propagate between {\em immediately 
separated} or {\em contiguous} events. In other words, in a 
finitary context the dynamical field of locality must somehow 
connect {\em nearest neighbouring events}, that is to say, events 
that define null Alexandrov intervals $A(x,y)=\emptyset$. Indeed, 
in \cite{rapt00a}, and based on an argument given in \cite{bpz}, it 
has been held that what is in fact subject to quantisation and, 
concomitantly, dynamical variability, is the {\em generating} or 
{\em covering} relation of the classical causal sets of Sorkin 
{\itshape et al.} Equivalently stated, only a reduction $\rota$ of 
the causally interpreted partial order $\to$ of the classical 
causal sets should be regarded as a quantum observable in the 
reticular-algebraic realm of quantum causal sets.    

\paragraph{Quantum causality and sheaves.} The significance of this 
assumption (the reduction of "$\to$" to "$\rota$") should not be 
underestimated.  For one thing, it lies at the heart of the 
sheaf-theoretic approach to {\em curved quantum causality} 
\cite{mallrapt00} as well as to the definition of a {\em 
non-commutative topology} for the latter \cite{rapt00c}.  This, 
without digressing much from our original theme, may be briefly 
described as follows: in section \ref{sclcorr} we noted the 
(anti)functorial equivalence between Sorkin's finitary poset 
substitutes of continuous spacetime manifolds (or their equivalent 
simplicial complexes) and their associated incidence algebras 
\cite{iasc}). In \cite{rapt00b} it was first observed that the 
contravariant functor relating the poset category of finitary 
substitutes of Sorkin with the poset category of their associated 
incidence Rota algebras is, in fact, an example of a {\em presheaf} 
\cite{buttish}. Subsequent {\em sheafification} of such presheaves 
gave rise to so-called {\em finitary spacetime sheaves}---algebraic 
spaces that may be interpreted as sound `coarse' approximations of 
the algebras of continuous ({\idest}, $C^{0}$) spacetime 
observables. 

The key structural feature of finitary spacetime 
sheaves is their (categorical) definition as {\em local 
homeomorphisms} between finitary topological posets and their 
associated incidence algebras, or in a causal context, between 
classical and quantum causal sets. Indeed, it has been extensively 
argued that the kinematics of quantum causal sets---which kinematics, 
in turn, stands for a locally finite, causal and quantal version of the 
kinematics of Lorentzian 
gravity\footnote{We tacitly assume that the kinematical structure 
suitable for classical Lorentzian gravity 
is that of a principal fiber bundle over a Lorentzian 
spacetime manifold, with structure group $SL(2,\cfield)$, and an $sl(2,\cfield)$-valued connection 
$\mathcal{D}$ on it.}---may be successfully modelled after 
non-trivial spin-Lorentzian connections defined on principal finitary 
spacetime sheaves of them and of their local gauge symmetry 
groups \cite{mallrapt00,rapt00c}. 
In turn, these connections may be thought of as encoding complete 
information about the dynamical variability of the local quantum 
causality relations $\rota$ between the elements of the objects in 
the respective categories, as well as they support a purely categorical 
version of the principle of general covariance of general 
relativity since the connections employed are indeed {\em finitary 
spacetime sheaf morphisms} \cite{mallrapt00,rapt00c}. In turn, this 
may be regarded as a significant preliminary step towards and 
entirely algebraic (in fact, categorical!) description of quantum 
gravity \cite{einst56,crane}.  

\paragraph{Comparison with Wheeler's spacetime foam.} We conclude 
this section by pointing out a fundamental difference between our 
approach to quantum causal foam and the `usual' conception of 
spacetime foam as originally conceived by Wheeler \cite{wheeler64}. 
Apart from the aforementioned important difference between the 
non-relativistic, undirected and `spatial' connotations of the 
usual mathematical term `topology' and the relativistic, directed 
and `temporal' undertones of the physical notion of `causality', 
topology is usually understood as the mathematical description of 
the global, `large scale' or gross features of space proper. 
Characteristically, Wheeler's definition of foam was as {\em 
dynamical variations and quantum fluctuations of global topological 
features of spacetime} (that is, spacetimes differing in 
holes, handles {\itshape etc}). That means, for instance, that 
amplitudes could be assigned to transitions between different 
topologies of `spacetime-as-a-whole'. 

In contrast, here we are more interested in {\em dynamics and 
quantum interferences of microlocality or microlocal causality}, 
since we are supposed to dissect spacetime into its finest causal 
elements at Planck scales---a realm where the classical notion of 
differential locality of the spacetime manifold is suspected to be 
grossly inadequate {\itshape vis-a-vis} quantum effects, let alone 
its global continuity (topological) properties. It is perhaps more 
appropriate altogether to describe our scheme as being 

\begin{itemize}
\item neither local---since locality inappropriately evokes ideas 
from the classical differential manifold model for spacetime, 
\item nor non-local---since we have not claimed yet any 
significant relevance of our models to the by now well established 
ideas on quantum non-locality, 
\end{itemize}

\noindent but {\dff alocal} \cite{qst}. After all, quantum causal 
sets still respect a novel quantum sort of locality that is by no 
means dependent on the pathological background geometric smooth 
spacetime continuum \cite{rapt00a,rapt00b,mallrapt00,rapt00c}. 
Perhaps it is in this new conception of a `{\em quantum locality}' 
that lies the strength and the value of the sheaf-theoretic 
approach to curved quantum causality, for one is actually able to 
stitch appropriately the `local experimental data' or `local 
spacetime measurements' ({\idest}, local sections of the 
curved finitary spacetime sheaves of quantum causal sets) and build 
the entire spacetime (sheaf) structure---topological, differential, 
indefinite metric {\itshape etc} from such local information 
\cite{mall,mallpriv}. Thus sheaf theory seems to provide us with 
powerful mathematical tools for implementing Bombelli {\itshape et 
al.}'s \cite{bomb87} deep insight that ``{\itshape causality as a 
partial order determines the topological, differential and metric 
structure of a Lorentzian spacetime manifold}''---one could say, 
{\itshape the raison d'\^etre} of the causal set approach to 
quantum gravity (see also \cite{sork90a}). 

Of course, as it was mentioned earlier, there still is some sense 
of `globalness' into our theoresis of the topology of spacetime as 
being a quantum observable, since we fundamentally assume that the 
results of measuring the spacetime topology are not numbers 
{\itshape per se}, but `entire' topological spaces---elements of 
the `topology spectrum' of spacetime foam as it was described in section 
\ref{salgmod}.

\section{Complementarity and the classical limit}\label{sclcorr}

The principal aim of this section is to present a way in which the 
algebras employed to model quantum causal foam may be thought of as 
encoding in their very structure some elements which may then be 
interpreted as manifestations of a kind of quantum indeterminacy. 
We will also discuss a mechanism by which the classical spacetime 
manifold of macroscopic experience and its local differential 
structure can emerge from such alocal algebraic quantum spacetime 
foam substrata. 

Since we take it as a fundamental issue that {\em spacetime itself 
is a quantum system} \cite{qst} whose main property, its (local) 
causal topology, is a quantum observable, we will argue how the two 
quintessential concepts of the usual quantum theory of matter, 
namely, Bohr's {\em complementarity} and the {\em correspondence 
principle} can be formulated in an entirely algebraic and 
combinatorial spacetime setting like ours. 

\subsection{Algebraic evidence for the uncertainty principle} 

Our elaborations with complementarity will depict a 
finitary-algebraic analogue of the indeterminacy relations between 
the so-called complementary or conjugate position/momentum 
observables of matter quanta which are usually thought of as 
propagating on a classical spacetime continuum. That is, since both 
the $q$ and $p$ observables of the usual (relativistic) matter 
quanta are most often assumed to have a continuous spectrum, the 
(locally) Euclidean manifold $\rfield^{4}$ model for spacetime 
appears to be almost inevitable \cite{buttish}.

The idea that a version of quantum complementarity is built into 
our finitary algebraic structures---an `innate indeterminacy' of 
our algebraic models of spacetime foam so to speak, may be briefly 
presented as follows: first we may recall from \cite{rapt00a} that 
with every classical causal set $\causet$ one may associate its 
incidence Rota algebra (quantum causal set, see section 
\ref{sclasscaus}) $\qauset$ in a way completely analogous to how we 
associated it with a finitary topological space in the beginning of 
section \ref{ssrotalg}. To do it, recall that any partial order on 
a finite set, in particular, the causal one, determines a finite 
topology, called {\em the order topology}. The convergence in this 
topology is just the order relation on the set $\causet$. 

\begin{equation}\label{elinqauset}
\causet=
\{x\to y\}
\quad\dashrightarrow\quad
\qauset(\causet)=
\lspan\{\ketbra{x}{y}:\, x\to y\in\causet\} 
\end{equation}

\noindent Now, if we consider the events of $\causet$ as some kind 
of configuration space for quantum causality, the Rota algebra 
$\qauset$ acts on the linear span of the events of $\causet$. The 
operators of the form $\ketbra{x}{x}$ may then be referred to as 
{\em quantum points-events}---they are the `causal self-incidences' 
in the quantum causal set $\qauset$ and, evidently, they correspond 
to the reflexivity relations in the underlying causal set 
$\causet$.  

Then, the `points' in the quantum causal set $\qauset(\causet)$ on 
which we impose the Rota topology, now the latter having a quantum 
causal interpretation \cite{rapt00a}, {\itshape \`a la} 
\ref{ssrotatop} are the (primitive) ideals in $\qauset(\causet)$ 
({\idest}, in our case, the kernels of equivalence classes of 
IRs of $\qauset(\causet)$) of the form:

\begin{equation}
I_{x}=\lspan\{\ketbra{y}{z}:~\ketbra{y}{z}\not=\ketbra{x}{x}\}
\end{equation}

\noindent each which can be decomposed into a sum 
$I_x=I_x^P+I_x^M$ with 

\[
\begin{array}{rcl}
I_x^P &=& 
\lspan\{\ketbra{y}{z}:~\ketbra{x}{x}\ketbra{y}{z}=0\}
\vphantom{\frac{\sum_a^b}{\sum_a^b}}\cr
I_x^M &=&
\lspan\{\ketbra{y}{z}:~\ketbra{x}{x}\ketbra{y}{z}\not=0\}
\end{array}
\]

For any $x$ all the elements of $I_x^P$ lie in the commutative  
subalgebra $\qauset^{0}:=\lspan\{\ketbra{x}{x}\}$ of $\qauset$. We 
regard them as operations of determination of {\em pure `position' 
states} of events of discretised quantum spacetime \cite{qst}.  
Effectively, $\qauset^{0}$ is the reticular and quantal analogue of 
the commutative algebra of coordinates or `position' states of the 
point-events of the classical spacetime manifold.

Furthemore, the operational quantum physical interpretation given 
to the elements of the  the linear subspaces $\qauset^{i}$ 
($i\geq1$) of $\qauset$ as operations of determinations of {\em 
pure `momentum' states} of events of discretized quantum spacetime 
concords with the fact that any pair of operations from $I_x^P$ and 
$I_x^M$ either have zero product, or do not commute. As for the 
latter there are precise algebraic Heisenberg commutation, the 
so-called `indeterminacy', relations that mathematically model 
them, also in the case of quantum causal sets there are abstract 
commutation relations that may be thought of as modelling the 
abstract complementarity described above. They are of the following 
kind:

\begin{equation}
[I_{x},I_{y}]:=I_{x}\cdot I_{y}-I_{y}\cdot I_{x}\not=0
\end{equation}

\noindent from which we may read the following `innate 
uncertainty':

\begin{quote} 
What constitutes a `quantum point-event' $I_{x}$ in a quantum 
causal set consists of all quantum spacetime momentum and path 
actions $\ketbra{y}{z}$ that exclude or `non-determine' the very 
position determination $\ketbra{x}{x}$ of the event-vertex $x$ in 
the underlying causal set $\causet$.  
\end{quote}

This is an abstract expression of the usual $q$--$p$ 
complementarity relations for matter quanta propagating in a 
continuous spacetime, and on them the whole edifice of a {\em 
non-commutative topology for curved quantum causality} rests, since 
indeed the product ideal $I_{x}\cdot I_{y}$ enters in a fundamental 
way into the very definition of the generating relation $\rota$ of 
the Rota (quantum causal) topology (described in \ref{ssrotatop}), 
now having a directly causal meaning as {\em immediate quantum 
causality} \cite{rapt00a,mallrapt00,rapt00c}. Furthermore, and here 
lies the mathematical versatility of our finitary algebraic models, 
it is our recognition of the $I_{x}$s as standing for abstract 
quantum point-events that makes it altogether possible to define a 
non-commutative Rota causal topology in a mathematically rigorous 
and semantically sound way \cite{mulpel1,rapt00c,mulpel2}.

\subsection{The emergence of classicality: correspondence 
principle} 

Our labours about the correspondence principle will essentially 
manifest how the classical spacetime of `macroscopic 
experience'\footnote{The inverted commas indicating that {\em 
actually we have no experience of a spacetime continuum}.}, with 
the commutative coordinate algebras of its events and the cotangent 
spaces of differential forms soldered on the latter, may be thought 
of as arising at the ideal limit of infinite localisation of its 
point-events (or at the experimentally implausible or 
`non-pragmatic' infinite power of resolution of spacetime into its 
point-events) \cite{sork91,qst,rapt00b}.

\paragraph{Functoriality.} As we have already mentioned in the end 
of section \ref{ssrotatop}, an arbitrary continuous mapping $\pkk$ 
between two finite topological spaces $\kkk'$ and $\kkk$ in general 
produces no homomorphism of their incidence algebras. However, if 
we restrict ourselves to spaces having the structure of simplicial 
complex, the situation becomes completely different. Namely, the 
correspondence $\kkk\mapsto\iak$ is a contravariant functor from 
the category of simplicial complexes to the category of 
differential modules over finite-dimensional commutative algebras.  
This dual mapping was first introduced in \cite{bdmh}, and the 
functorial properties were studied in detail in \cite{iasc}. It 
differs from the Eilenberg--Steenrod duality \cite{eilsteen}---cf. 
\eqref{e62}. 

In \cite{sork91} inverse systems of finitary substitutes were 
considered which made it possible to reconstruct (or to have as a 
limit) a $C^{0}$-topological manifold space. Similarly, and from 
the categorically dual point of view of (algebras of) continuous 
({\idest}, $C^{0}$) functions on the topological base 
spacetime manifold---the so-called `algebras of continuous 
spacetime observables', in \cite{rapt00b} inverse systems of 
finitary spacetime sheaves of continuous functions were also seen 
to `converge' to the $C^{0}$-sheaf of continuous observables as the 
`microscopic' power of resolution of spacetime into its 
point-events increased without bound. At this purely topological 
level there was no way one could introduce differential structures 
in intermediate finitary steps.  Now, having this functoriality, we 
can consider direct limits of the Rota algebras of topological 
spaces. They are graded algebras, and the morphisms preserve 
grading as well as they intertwine with the Cartan-K\"ahler 
differential. This gives us the opportunity to treat intermediate 
finite spaces as full fledged spacetime substitutes, capturing 
both the topology ({\idest}, `convergeability') and the 
differential geometric structure ({\idest}, 
`differentiability'). 

When spacetimes are subsituted by finitary topological spaces, we 
may consider finer or coarser experiments. That is why we have to 
formalize the notion of refined experiment. Within the Sorkin 
discretisation procedure described in \ref{ss21} a refinement means 
passing to an inscribed covering of the manifold. In this case any 
element of the finer covering is contained in an element of the 
coarser one.

\paragraph{Construction of classical limit.} Each step of a 
limiting procedure, that is, of a refined covering, gives rise to a 
projection of appropriate complexes: the finer one is projected to 
the coarser one. In the dual framework we have an injection of the 
smaller algebra associated with a coarser measurement to a bigger 
one associated with a finer one.

In general, limiting procedures for approximating systems (whatever
these may be, posets or algebras) are organised using the notion of
converging nets. Namely, each pragmatic observation is labelled by
an index $\alpha$ and we have the relation of refinement $\succ$ on
observations: $\alpha \succ \alpha'$ means that the observation
$\alpha$ is a refinement of $\alpha'$.

When we are dealing with posets with each pair $\alpha,\alpha'$
such that $\alpha \succ \alpha'$ a canonical projection
$\om_{\alpha'}^\alpha : \kkk_\alpha \to \kkk_{\alpha'}$ is defined
such that for any $\alpha \succ \alpha' \succ \alpha''$ the 
following coherence condition holds: 

\[
\om_{\alpha''}^\alpha =
\om_{\alpha''}^{\alpha'}\om_{\alpha'}^\alpha
\]

\noindent The functoriality requirement is evoked when 

\begin{equation}\label{e77}
\om(\kkk^0_\alpha) \subseteq \kkk^0_{\alpha'}
\end{equation}

\medskip 

Then we introduce the set of threads. A {\dff thread} is a
collection $\{t_\alpha\}$ of elements $t_\alpha \in \kkk_\alpha$
such that

\[
t_{\alpha} = \om_{\alpha}^{\alpha'} t_{\alpha'}
\]

\noindent whenever $\alpha \succ \alpha'$. Denote by $\tttt$ the
set of all threads.

The next step is to make $\tttt$ a topological space which is done 
in a standard way \cite{al56}: $\tttt$ is a subspace of the total 
cartesian product $\tttt_0 = \times_\alpha \kkk_\alpha$. Endow 
$\tttt_0$ with the product Tikhonov topology, then $\tttt$ being a 
subset of $\tttt_0$ becomes a topological space. Finally, we obtain 
the limit space $X$ as the collection of all closed threads from 
$\tttt$.  This procedure is described in detail in \cite{sork91}. 

A detailed scheme for building the dual limit---that of algebras 
was presented in \cite{landi99}. As mentioned above, with any pair 
$\alpha\succ\alpha'$ of pragmatic observations we have a canonical 
injection

\[
\om^{*\,\alpha'}_{\alpha} : \ia(\kkk_{\alpha'}) \to
\ia(\kkk_{\alpha})
\]

\noindent Moreover, due to the requirement \eqref{e77} the 
restriction of each $\om^*$ on commutative subalgebras 
$\aaa=\ia^0\subseteq\ia$ is well defined. Now we first consider the 
set of all sequences

\[
{\mathbf{\Omega}} = \times_\alpha \ia(\kkk_\alpha) =
\{ \{a_\alpha\}\,|\, a_\alpha \in \ia(\kkk_\alpha) \}
\]

\noindent and select the set of {\em converging} sequences in the
following way. Note that $\aaa$ is an algebra. Introduce a norm
$\lVert\cdot\rVert_\alpha$ in each finite-dimensional algebra
$\ia(\kkk_\alpha)$, then a sequence $\{a_\alpha\}$ converges if
and only if for any $\epsilon > 0$ there exists a filter
$\fff_\epsilon$ of indices $\alpha$ such that

\[
\forall \alpha, \alpha'\in \fff_\epsilon
\quad \alpha \succ \alpha' \Rightarrow
\lVert\om^{*\,\alpha}_{\alpha'}a_\alpha - 
a_{\alpha'}\rVert_{\alpha'} < 
\epsilon 
\]

Since any element of the limit algebra is a net, we can consider the
coupling between the limit algebra and the limit space which
consists of nets. The result of this coupling is a converging net
of numbers whose limit may be thought of as the value of an element of
the limit algebra at a point of the limit space. At present the 
rigorous results are obtained and verified only for the functorial 
part of the construction, for details the readers are referred to 
\cite{iasc}. 

\paragraph{Classicality in the limit.} The conclusion of the last 
paragraph prompts us to first discuss briefly 
the mathematical character of the `classical limit' or `end space' alluded to 
above, before delving into the physical interpretation proper of the 
aforedescribed inverse or projective limit process on the 
inverse system or net of Rota algebras as well as into the physical  
nature of the end space that this procedure yields.

Perhaps the most important characteristic of the inverse limit 
space is that it can qualify as a {\em manifold} proper.  We have 
already mentioned in section \ref{salgmod} in connection with 
Sorkin's poset discretisations of topological manifolds in 
\cite{sork91} and in the beginning of this subsection that 
Rota algebras are objects (categorically) dual to those finitary 
posets. In the particular case that the latter are also (assumed to 
be) locally finite simplicial complex `nerve-skeletonisations' of 
$C^{0}$-manifolds {\itshape \`a la} \v{C}ech-Aleksandrov 
\cite{eilsteen, al56} as it is particularly of interest to us 
here\footnote{These are the `good' posets alluded to in subsection 
1.4.}, the categorical duality between finitary posets/simplicial 
complexes and Rota algebras secures that, as an inverse system of 
finer-and-finer finitary poset topologies `converges' to a space 
({\itshape ie}, possesses an inverse limit space) that for all 
practical purposes is homeomorphic to ({\itshape ie}, topologically 
indistinguishable from) a topological manifold \cite{sork91, 
rapt00b}, so a projective system of their dual Rota algebras would 
also yield a $C^{0}$-manifold-like structure\footnote{After 
all, it would be begging the question and it would put the terms 
`approximations', `substitutes' or `replacements' to doubt, if one 
started with a topological manifold structure that one would like 
to approximate by finitary means---be it finitary posets as in 
\cite{sork91} or their dual Rota algebras as here---and finally  
({\itshape ie}, after the inverse limit procedure) one would 
recover a non-manifold topology!}.

Moreover, the bonus in our case is that the Rota algebras involved, 
being discrete differential spaces, encode information not only 
about the topology of the continuous space that they are supposed 
to substitute (and recover at a suitable projective limit), but also about its 
differential (or differentiable) attributes. In case we 
substitute a smooth manifold, one would then have strong reasons to 
believe, as we do, that the aforementioned end space would 
{\itshape a fortiori} be a $C^{\infty}$-smooth manifold, not just a 
$C^{0}$-continuous one. On the other hand, it is admittedly conceivable that 
although every `approximating term' (Rota algebra) in the net may 
be `nice'\footnote{As we said in the previous footnote, for us, 
`niceness' means `the Rota algebra of a simplicial complex'; see 
section \ref{salgmod}.}, the end space may turn out to be quite 
pathological ({\itshape eg}, `fractal'), or even more surprisingly, 
that while the maps between the approximating spaces are relatively 
regular, the limit space comes out rather singular. Certainly, 
we should not 
fully commit ourselves to recovering a smooth continuum; after all, 
we should allow for `pathologies at the limit' for they would 
perhaps point at (potentially useful) features to be rectified at 
the finite level of the approximating spaces and their 
corresponding Rota algebras. Thus, perhaps it 
would be more appropriate instead of calling the Rota algebras 
involved `discrete differential manifolds'---the reticular 
or combinatorial analogues of (the cotangent spaces of differential forms on) the 
$C^{\infty}$-smooth manifold \cite{qst, mallrapt00, rapt00c}, to 
coin them `discrete differential spaces'---finitistic models of a 
differential geometry not necessarily identical to the usual 
calculus on smooth manifolds ({\itshape ie}, the standard 
differential geometry)\footnote{We would like to thank 
cordially the referee of Classical and Quantum Gravity for 
sensitising us to the possibility that apparently `healthy' 
finitary approximations may yield at the limit pathological 
differential continua supporting rather singular differential calculi.}.

More on the interpretational side, the suggested limit 
constructions play the crucial r\^ole of providing a firm (formal) 
algebraic background on which to (semantically) support the ever so 
important for the physical theory correspondence principle 
otherwise known as the method of making contact between the quantum 
and the classical models of the system in focus (here, spacetime).  
Thus, in closing this section, let us take the opportunity to 
recall from \cite{qst} and discuss briefly the physical 
interpretation given to these formal limit procedures as `classical 
limit processes'---processes of emergence of the classical 
spacetime manifold of macroscopic scales from a wildly fluctuating 
primordial `pool' (or `soup'!) of alocal quantum causal topological 
foam substrata (always keeping in mind however the 
possibility mentioned in the previous paragraph that the limit 
space may not be free from `differential abnormalities'). 

\paragraph{How classical is the classical limit?} In \cite{qst} 
what principally motivated us to interpret the limit processes on 
discretized quantum spacetime topologies as Bohr's correspondence 
principle was the cogent quantum interpretation that the underlying 
Rota algebraic structures afforded, their prominent discrete 
differential geometric characteristics, as well as typical traits 
of the classical $C^{\infty}$-smooth spacetime manifold that were 
already foreshadowed at the reticular level of our structures.  
Thus we envisaged a process of `decoherence' (or anyhow, of 
breaking or destruction of quantum coherence) of the topological 
arrow-connections in the alocal quantal (and here, causal) Rota 
algebras as our finitary experimenters employed higher power ({\itshape  
ie}, presumably higher microscopic energy) to resolve spacetime 
into its point-events, as it were, {\em to localise it maximally}. 

\medskip 

Being faithful to an operationalistic philosophy for 
physics and respecting the main lesson we have learned from 
applying quantum theory to spacetime structure ({\idest}, that one 
cannot actually localize the gravitational field below Planck 
length without creating a singularity in the underlying spacetime 
manifold structure), we gathered that such maximal spacetime 
localisations are ideal (mathematical) constructions of 
non-pragmatic physical significance\footnote{Ever ready to 
reconsider however if and when experimentalists show us that one 
can actually record a real number in the laboratory.} justified 
only by the fact that the model for spacetime at large scales that 
works best (namely, the differential continuum) can be obtained at 
such a limit of infinite refinement of the (causal) foam pool.  
Indeed, since the local structure of {\em classical} spacetime 
({\idest}, the commutative coordinate algebras of its point-events 
and the spaces of differential forms cotangent to the latter) is 
recovered at such an infinite localisation limit of the granular 
and quantal foam algebras, our identification of it with the usual 
correspondence principle (or limit) due to Bohr is well justified 
even if {\itshape a posteriori} it seems that our original 
motivation was quite `ophelimistic' ({\idest}, that the 
end-manifold somehow justifies the means-algebras exactly because 
of the existence of the mathematical limit constructions linking 
the two). 

Altogether however, a `die-hard' physicist may object to our 
`emergence of classicality' scheme above by claiming that the 
classical spacetime manifold is the result of an {\em actual 
physical process} such as condensation for instance \cite{df88, 
df96, bfsz}, so that a `purely kinematical' (or structural) account 
of `{\em how come the continuum?}' should give way to a more 
dynamical scenario. At the same time, a `quantum purist' may also object 
to the way we use Bohr's correspondence principle in that, in quite a 
general sense, Bohr intended to use this notion in close connection 
with the process of measurement and the resulting recording of 
c-numbers from q-systems; while, strictly speaking, we argue here 
that, in fact, we measure or `extract' entire topological spaces 
from quantum spacetime (see sections \ref{salgmod} and 
\ref{sclasscaus}), not merely commutative numbers. Granted that we 
recognise that our algebraic description of spacetime 
foam has been characterised as being more `kinematical' than 
`dynamical'\footnote{See Afterword in this 
paper, as well as \cite{mallrapt00, rapt00c}.} and that we declared 
up-front that we regard the spacetime's (causal) topology as being 
a quantum observable, such criticism, although reasonably justified, 
could hardly present any significant hindrance to the 
further development of the theory of quantum spacetime structure 
propounded herein.
   
\section{A toy model}\label{stoymodel}

In this section we present an explicit example of 49-dimensional
Hilbert space endowed with a product operation, and a self-adjoint
operator in this space such that for two its different eigenspaces
the spatialisation procedure yields two topologically different
spaces: a piece of plane and a circle considered in the example 
above.

Let $\hhh$ be the space of all complex valued $7\times 7$
matrices:

\[ 
\hhh = M_7 = 
\lspan\{\eee_{ik}\mid i,k=1,\ldots, 7\}
\]

\noindent choosing matrix units introduced above in \eqref{eab} as its 
orthonormal basis: 

\[ 
(\eee_{ik},\eee_{i'k'}) = 
\delta_{ii'}\delta_{kk'}
\] 

The product operator on $\hhh$ which is necessary to define the 
Rota topology will be the usual matrix product:  
$\eee_{ik}\cdot\eee_{i'k'} = \delta_{ki'}\eee_{ik'}$. Denote by 
$P_{ik}:\hhh\to\hhh$ the projector on the basis element 
$\eee_{ik}$ and, denoting by $\sfproj(a)$ the orthogonal projector 
on a unit vector $a$, define the following projectors:

\[
\begin{array}{rl}
P_{c2} &= \sfproj\left(\
\frac{1}{\sqrt{2}}
(\eee_{62}+\eee_{72})
\right)\cr
P_{c3} &= 
\sfproj\left(
\frac{1}{\sqrt{2}}
(\eee_{63}+\eee_{73})
\right)\\
P_{pc} &= 
\sfproj\left(
\frac{1}{\sqrt{179}}
(\eee_{11}-2\eee_{22}+3\eee_{33}-4\eee_{44}
+6\eee_{55}+7\eee_{66}-8\eee_{77})
\right)\\
P_{cc} &= 
\sfproj\left(
\frac{1}{\sqrt{95}}
(6\eee_{11}+5\eee_{22}+4\eee_{33}+3\eee_{44}
+\eee_{55}+2\eee_{66}+2\eee_{77})
\right)
\end{array}
\]

Consider the self adjoint operator $\ttt:\hhh\to\hhh$:

\begin{equation}\label{etoyoper}
\begin{array}{ll}
\ttt &=
(P_{14}+P_{15}+P_{25}+P_{26}+P_{36}+P_{34}+P_{47}+P_{57}+P_{67}+P_{pc})-\cr
&-(P_{41}+P_{43}+P_{51}+P_{52}+P_{c2}+P_{c3}+P_{cc})
\end{array} 
\end{equation}

\noindent having three eigenvalues: 0, +1 and -1. 

\medskip 

If we carry out the measurement associated with $\ttt$, the result 
will be 0, 1 or -1. If the result is +1, then we are in the 
subspace of $\hhh$ generated by the first line of \eqref{etoyoper}.  
The algebra spanned on this subspace is isomorphic to the matrix 
algebra $\ia_{\hbox{\small plane}}$ from \eqref{exofalg}. When the 
eigenvalue is -1, we get the eigenspace whose algebraic closure is 
isomorphic to $\ia_{\hbox{\small circle}}$ from \eqref{exofalg}. 
When the eigenvalue is 0, the span of the appropriate eigenspace is 
the full matrix algebra, and the resulting space consists of one 
point. 

\medskip 

To clarify the constructions of this section let us 
consider in more detail what happens if we obtain the eigenspace 
associated with $+1$ eigenvalue. Begin with its algebraic closure. 
All the functions of the vector 
$\eee_{11}-2\eee_{22}+3\eee_{33}-4\eee_{44}+ 
6\eee_{55}+7\eee_{66}-8\eee_{77}$ will form the commutative algebra 
${\rm Diag}_7$ of all diagonal $7\times 7$ matrices:

\[ 
{\rm Diag}_7=
\lspan_{\cfield}\nolimits\left\{
\eee_{11},\eee_{22},\eee_{33},\eee_{44},
\eee_{55},\eee_{66},\eee_{77}
\right\}
\] 

\noindent Then recall that according to \eqref{eab} we have

\[
\begin{array}{rl}
\eee_{14}\eee_{47} &= \eee_{17}\cr
\eee_{25}\eee_{57} &= \eee_{27}\cr
\eee_{36}\eee_{67} &= \eee_{37}\cr
\end{array}
\]

\noindent and so we get the algebra $\ia_{\hbox{\small plane}}$ 
from \eqref{exofalg}. It is graded:  

\begin{equation}\label{etoybasis}
\begin{array}{ll}
\ia^0 &=\lspan_{\cfield}\left\{
\eee_{11},\eee_{22},\eee_{33},\eee_{44},
\eee_{55},\eee_{66},\eee_{77}
\right\}\cr
\ia^1 &=\lspan_{\cfield}\{
\eee_{14},\eee_{15},\eee_{25},\eee_{26},\eee_{36},
\eee_{34},\eee_{47},\eee_{57},\eee_{67}\}\cr
\ia^2 &=\lspan_{\cfield}\{
\eee_{17},\eee_{27},\eee_{37}\}
\end{array}
\end{equation}

\medskip 

The total algebra $\ia$ is 

\[
\ia=\ia^0\oplus\ia^1\oplus\ia^2
\]

\noindent It can be described by the template \eqref{exofalg} 
(the right matrix). Let us spatialise it according to the recipe of 
section \ref{ssrotatop}. There are 7 IRs of $\ia$. To specify 
them, denote by $a=\sum_{ik}a_{ik}\eee_{ik}$ where the pairs $ik$ 
range over the values occurring in \eqref{etoybasis}. Then the IRs 
of $\ia$ are: 

\[
\begin{array}{l@{$\quad$}l}
\mathbf{1}:a\mapsto a_{11}&\mathbf{5}:a\mapsto a_{55}\cr  
\mathbf{2}:a\mapsto a_{22}&\mathbf{6}:a\mapsto a_{66}\cr  
\mathbf{3}:a\mapsto a_{33}&\mathbf{7}:a\mapsto a_{77}\cr  
\mathbf{4}:a\mapsto a_{44}&
\end{array}
\]

\noindent Now let us describe the Rota topology on the set of the 
points (=IRs) that we obtained. First calculate the relation $\rota$ 
\eqref{edefrota}. Take two representations, say, 1 and 4, and 
consider their kernels $x_1$ and $x_4$, and then their intersection 
$x_1\cap x_4$ and the product $x_1\cdot x_4$: 

\[
\begin{array}{rcl}
x_1&=&\left\{
\sum_{ik}a_{ik}\eee_{ik}
\quad\mbox{such that}\quad
ik\neq11\right\}\cr
x_4&=&\left\{
\sum_{ik}a_{ik}\eee_{ik}
\quad\mbox{such that}\quad
ik\neq44\right\}\cr
x_1\cap x_4&=&\left\{
\sum_{ik}a_{ik}\eee_{ik}
\quad\mbox{such that}\quad
ik\neq11,\,44\right\}\cr
x_1\cdot x_4&=&\left\{
\sum_{ik}a_{ik}\eee_{ik}
\quad\mbox{such that}\quad
ik\neq11,\,44,\,14\right\}
\end{array}
\]

\noindent and we see that $x_1\cap x_4\neq x_1\cdot x_4$, therefore 
according to \eqref{edefrota} $x_1\rota x_4$. Checking in the same 
way that $xy\neq x\cap y$ \eqref{edefrota} for all pairs $i,k$ we get: 

\[
\begin{array}{c@{$\qquad\quad$}c@{$\qquad\quad$}c}
1\rota4 & 1\rota5 & 2\rota5\\
2\rota6 & 3\rota4 & 3\rota6\\ 
4\rota7& 5\rota7& 6\rota7 
\end{array}
\]

\noindent Forming the transitive closure \eqref{eqxy} of the 
relation $\rota$, we finally obtain the convergence graph drawn at 
figure \ref{fcircle} a). So, the topological space is 
reconstructed, let us study its differential properties.

\medskip 

The basic algebra $\aaa=\ia^0$ is commutative, and the space of 
`covectors' $\ia^1$ possesses the structure of a (say, left) 
$\aaa$-module with the action of $\aaa$ defined by the 
multiplication \eqref{eab}. This `cotangent space' is 
2-dimensional, namely, any element of $\ia^1$ can be generated 
acting by the algebra $\aaa$ on the following two elements: 

\[
\begin{array}{l}
\sfvec_1=\eee_{14}+\eee_{25}+\eee_{36}+
\eee_{47}+\eee_{57}+\eee_{67}\cr
\sfvec_2=\eee_{15}+\eee_{26}+\eee_{34}
\end{array}
\]

\noindent that is, 
$\ia^1=\lspan_\aaa\nolimits\{\sfvec_1,\sfvec_2\}$. 

\noindent Similarly, the space of 2-forms is one-dimensional, its 
only generator is the element 
$\sfvec=\eee_{17}+\eee_{27}+\eee_{37}$, so 
$\ia^2=\lspan_\aaa\nolimits\{\sfvec\}$. So, we see that the 
differential geometry of the substitute is really similar to that 
of a piece of a plane. 

\medskip 

A similar construction can be carried out for the eigenspace of 
$\ttt$ associated with the value -1. We remark that in this case 
the algebra $\aaa$ generated by the functions of the element 
$6\eee_{11}+5\eee_{22}+4\eee_{33}+3\eee_{44} 
+\eee_{55}+2\eee_{66}+2\eee_{77}$ will have dimension 6 rather than 
7, namely: 

\[ 
\aaa = 
\lspan_\cfield\nolimits\{
\eee_{11},\eee_{22},\eee_{33},\eee_{44},
\eee_{55},\eee_{66}+\eee_{77}\}
\simeq {\rm Diag}_6
\] 

\noindent then, if we proceed in a similar way, we obtain the finitary 
substitute of a circle whose topology is depicted in figure 
\ref{fcircle} and whose Rota algebra has the template matrix 
\eqref{exofalg}.

\section*{Afterword}

We conclude the present paper by commenting on six issues that one 
may legitimately bring up as critiques of the approach to spacetime 
foam presented above.

\medskip 

First, if the theoretical scheme ultimately aspires to be dynamical 
and at the same time alocal, one may reasonably ask: {\em dynamics 
of quantum causal topology with respect to what?} However, if there 
is such an external parameter space, or more colloquially, a 
background clock relative to which to refer the dynamical changes 
in quantum causality, is not this another manifestation of an 
ether-like absolute substance---another unobservable of the theory 
that is of dubious physical relevance? Our response to this is that 
each experimenter or observer of quantum causality creates her own 
picture of the dynamics of quantum causality; as it may, she 
creates her own `quantum wristwatch' or `quantum causal or time 
gauge', so that it is understood that a dynamics of quantum 
causality must be manifestly invariant or `causal gauge covariant' 
relative to these gauges or `local measurements' 
\cite{mallrapt00,rapt00c}.

\medskip

Second, we must admit that as yet we have not been able to arrive 
at a full fledged dynamical scheme for spacetime foam and, 
{\itshape in extenso}, for quantum causality, so that it is perhaps 
more accurate to call our approach {\em the kinematics of quantum 
causal topology} \cite{rapt00a,mallrapt00,rapt00c}.  For the 
closely related causal sets and their ultimate aim to model quantum 
gravity \cite{bomb87,sork90a,sork90b,ridesork00}, Sorkin has 
convincingly argued for the importance of first acquiring a firm 
grasp of the kinematics of a theory before tackling the problem of 
dynamics \cite{sork95}. Overall, and having in mind the classical 
paradigm of the development first of the special and then of the 
general theory of relativity in that the former defined the 
kinematical structure of causality ({\idest}, the Minkowski 
lightcone soldered at each event) while the latter formulated its 
localisation or gauging and concomitant curving ({\idest}, the 
dynamics of locality as encoded in $g_{\mu\nu}(x)$), it seems more 
reasonable to us to formulate first how quantum spacetime topology 
can {\em possibly} move ({\em ie}, its kinematical structure), and 
then entertain ideas of how it {\em actually} moves ({\em ie}, its 
dynamics). In this light, the correspondence limit manifold 
structure for spacetime that we are going to give should be 
properly called {\em a kinematical mechanism for recovering the 
classical spacetime continuum}\footnote{We wish to thank Chris 
Isham and Lee Smolin for pointing this out in private 
communication.  See also \cite{mallrapt00,rapt00c}.}.

\medskip 

Third, the use of non-$*$-algebras is often claimed to be 
unphysical. Related to the discussion above about kinematics {\em 
versus} dynamics is Penrose's \cite{penr87} suggestion that ``the 
true quantum gravity is a time asymmetrical theory". We also expect 
our theory to address somehow the problem of the quantum arrow of 
time {\em vis-a-vis} dynamical quantum causality and the related 
conception of spacetime foam. To begin with, the algebras that we 
use are non-involutive ({\em ie}, they are not $*$-algebras) unlike 
the algebras of observables of relativistic matter quanta used in 
quantum field theory over the flat Minkowski space. However, these 
field theories, as well as the classical field theory of gravity 
({\em ie}, general relativity), are manifestly time reversal 
invariant so that physics down to Planck length and duration 
($l_{P}$-$t_{P}$) appears to be predominantly time symmetric with 
the outstanding exception of the chiral, parity violating weak 
interactions at energies orders of magnitude lower than $E_{P}$. 
From an operational point of view the non-commutative $*$-algebras 
of quantum matter field observables employ the unary involution 
operation to model reversal of time, or more generally, of the 
sequential order of operation, and they are closed under it. In 
contradistinction, our algebras are not closed under any sort of 
involution hence they give an early indication of an asymmetry 
built into their very structure {\em ab initio}.  Indeed, in an 
algebraic model of dynamical quantum causal topology closely 
related to the one presented above, it was argued \cite{rapt00c} 
that the fundamental quantum time asymmetry must be encoded already 
at the structural or kinematical level of the theory so that the 
dynamics merely `carries', respects or preserves this kinematical 
quantum arrow of time.

\medskip 

Fourth, it has also been widely suspected that in the quantum deep, 
precisely due to the potent dynamical fluctuations that the whole 
spacetime edifice is subject to, there should not be a sharply 
defined local causal structure.  The local lightcone of the 
classical curved spacetime manifold of general relativity becomes 
some kind of a fuzzy structure in the sense that the classical 
relativistic causal connections between events should be replaced 
by a more general, because probabilistic, sort of connections 
between quantum states of spacetime events.  Indeed, Finkelstein 
\cite{df69} in a discrete, relativistic and quantal model of 
spacetime structure and causality arrived at an irreducible 
uncertainty relation between the space and time coordinates---a 
relation that {\em a fortiori} points to the implausibility of 
talking about a clear-cut local lightcone structure in the quantum 
regime. Here too the algebras employed model quantum states of 
events and kinematical paths for possible dynamical transitions 
between various quantum spacetime topologies \cite{qst}. In a 
purely causal interpretation of these algebras \cite{rapt00a} we 
still hold that it is meaningful to speak of well defined causal 
connections between events, albeit, both the connections and (the 
states of) the events themselves are inextricably entangled by 
coherently superposing with each other.  In section 
\ref{sclasscaus} we show that there are quantum uncertainty 
relations built into our algebraic models of spacetime foam.

\medskip

The fifth criticism of our approach to spacetime foam is a general, 
rather conceptual one and it is based on the now famous advice that 
Einstein gave to Heisenberg, namely, that {\em it is the theory 
that determines what is observable} in the sense that no 
theoretical entities should be declared up-front or `postulated' to 
be observables \cite{heis96}. The boundary between what can be 
observed and what cannot is delimited by the theoretical model 
itself\footnote{Interestingly enough, in Greek `theory' 
(`$\theta\epsilon\omega\rho\acute{\iota}\alpha$') is almost 
tautosemous to `observation' 
(`$\pi\alpha\rho\alpha\tau\acute{\eta}\rho\eta\sigma\eta$')!}.  
Since by `theory' the physicist understands `dynamics', and since 
we have not proposed one for quantum causality as yet, our 
{\itshape a priori} intuitive assumption of a dynamically variable 
quantum causal topology is on a par with the primitive intuition of 
the early Einstein long before he arrived at the concrete dynamics 
of general relativity that the spacetime geometry is the sole 
dynamical variable in the theory of gravity ({\idest}, the 
components of the spacetime metric $g_{\mu\nu}$ were insightfully 
identified with the gravitational potentials) \cite{torr}. The 
aforementioned advice of Einstein to let the theory select the 
measurable dynamical quantities greatly surprised and impressed 
Heisenberg \cite{heis96}. 

\medskip

Finally, our strong commitment to finite dimensional algebras and 
their finite dimensional IRs throughout this paper may strike odd a 
thinker who is aware of one of the most serious obstacles in 
attempts at uniting relativity and quantum theory, namely, that 
there are no finite dimensional unitary IRs of the non-compact 
orthochronous Lorentz group. At this early stage of the 
construction of the theory it seems more natural to us to 
sacrifice unitarity for finiteness since the former is 
usually perceived as a non-local conception since (in 
non-relativistic quantum mechanics) it conventionally involves an 
integration over all space(time).  Also, the strong relationship 
between unitarity and the probabilistic ({\idest}, positive 
definite metric/measure) interpretation of quantum mechanics does 
not seem to accord with the indefinite signature of the spacetime 
metric generated by a causality modelled after a partial order 
\cite{robb,bomb87}, thus we are quite keen on abandoning unitarity 
altogether due to the more local and indefinite nature of our 
structures. To our knowledge, such an urge to study Hilbert spaces 
with an indefinite metric and their associated pseudo-unitary 
structure with its concomitant pseudo-probabilistic interpretation 
was prophetically anticipated in \cite{heis54,feyn76}. On the other 
hand, another option open to us is to keep some sort of pseudo-unitarity 
in the theory and try to regard as `effective symmetries' of our structures 
only those belonging to a finite subgroup of the Lorentz group---the 
transformations that are, say, `available' to and link our finitary 
experimenters. For in any case, it has been seriously proposed that 
since a Lorentzian spacetime manifold may be regarded as a 
structure `effective' only at classical scales, its (local) Lorentz 
symmetries should also be regarded as {\em a macroscopic quantum 
effect}---a quantum physically non-fundamental idiosyncracy of the 
classical continuum which itself is the result of a process of 
quantum condensation\cite{df96,bfsz}. The motivating 
insight for this proposal is that if the system (here, spacetime) 
is fundamentally taken to be quantum, then so must its symmetries 
\cite{df96}. Similarly, in \cite{mallrapt00} it has been proposed 
that since the differential Lorentzian manifold has been 
effectively `reticularized', so must be the group manifold of its 
local spin-Lorentz ({\idest}, $SL(2,\cfield)$) gauge 
symmetries.

In closing this paper we would like to speculate very briefly on 
the future development of this approach of ours to spacetime foam. 
For one thing, since we have not provided an explicit dynamics for 
quantum causal topology, we are still haunted by Sorkin's original 
question: `{\em How does one vary a (locally finite) poset?}' 
\cite{sorkpriv}, which, in the quantum causal set context, 
translates to: `{\em How does one vary a (finite dimensional) 
ring?}---the basic shortcoming of our scheme so far being that it 
provides a possible model of the kinematics of an observable 
quantum causality. Loosely speaking, what we know now is: {\em 
`Where does one vary a poset'}---the `kinematical rooms' in which 
quantum causality can possibly vary; while a concrete dynamics 
still remains an open theoretical question.  Evidently, this then 
seems to be the future of our quest: find a cogent dynamical 
scenario for quantum causal sets. Formally, significant advances 
towards this end have already been made from a sheaf and 
topos--theoretic point of view \cite{mallrapt00,rapt00c}, but still 
a lot of work has to be done in order to bring together diverse 
ideas from the by now quite wide causal set approach to quantum 
gravity and amalgamate them with our algebraic approach to 
spacetime foam described above.

\paragraph{Acknowledgments.} IR wishes to thank Chris Isham and 
Tasos Mallios for numerous exchanges on quantum causality and 
advice on its possible dynamics, as well as Chris Mulvey for 
helpful interactions on what `ought to' qualify as non-commutative 
topology proper.  He also acknowledges the financial support of the 
EU in the form of a Marie Curie 
Research Fellowship. RRZ acknowledges the `encouraging criticism' 
by A.A.Grib and S.Krasnikov and the financial support from the 
European Research Project Q-ACTA and from the state research 
program `Universities of Russia'. In the revised version of the 
paper the authors express their gratitude to the referee of 
`Classical and Quantum Gravity' for essential and highly 
constructive remarks.

\end{document}